\documentclass[twocolumn,superscriptaddress,secnumarabic,amssymb,nobibnotes,aps,prl,numerical]{revtex4-1}

\usepackage{graphicx}
\usepackage{verbatim}
\usepackage{mathrsfs}

\usepackage[normalem]{ulem}
\usepackage{color}

\begin{document}

\title{Non-equilibrium dynamics of coupled qubit-cavity arrays}


\author{Felix Nissen}
\address{Cavendish Laboratory, University of Cambridge, Cambridge CB3 0HE, United Kingdom}

\author{Sebastian Schmidt}
\address{Institute for Theoretical Physics, ETH Zurich, 8093 Zurich, Switzerland}

\author{Matteo Biondi}
\address{Institute for Theoretical Physics, ETH Zurich, 8093 Zurich, Switzerland}

\author{Gianni Blatter}
\address{Institute for Theoretical Physics, ETH Zurich, 8093 Zurich, Switzerland}

\author{Hakan E. T\"ureci}
\address{Department of Electrical Engineering, Princeton University, Princeton, New Jersey 08544, USA}
\address{Institute for Quantum Electronics, ETH-Z\"urich, CH-8093 Z\"urich, Switzerland}

\author{Jonathan Keeling}
\address{Scottish Universities Physics Alliance, School of Physics and Astronomy, University of St Andrews, St Andrews KY16 9SS, United Kingdom}

\begin{abstract}
We study the coherence and fluorescence properties  of the coherently pumped and dissipative Jaynes-Cummings-Hubbard model describing polaritons in a coupled-cavity array. At weak hopping we find strong signatures of photon blockade similar to single-cavity systems. 
At strong hopping the state of the photons in the array depends on its size. While the photon blockade persists in a dimer consisting of two coupled cavities, a coherent state forms on an extended lattice, which can be described in terms of a semi-classical model.
\end{abstract}
\maketitle

The realization of effective photon-photon interactions in various cavity QED systems has triggered an immense interest
in using these light-matter systems for quantum computation and simulation.
A central element of this work is the photon blockade \cite{blockade}, where the presence of a single photon in a driven cavity
prevents more photons from entering. Key experimental signatures of this effect are anti-resonant lineshapes in
homodyne/heterodyne detection \cite{bishop-nphys} as well as photon anti-bunching and the appearance of a dressed state Mollow triplet in resonance fluorescence spectra \cite{wallraff}.
So far, these signatures have been observed in single cavity systems.

With the unprecedented experimental control of recent experiments at hand, a key challenge in the study of coupled light-matter systems is the interplay of strong correlations and collective behavior in extended systems. Recent experimental progress includes 
the realization of a Tavis-Cummings nonlinearity with superconducting qubits in circuit QED \cite{Fink2009,Fink2008},  a BEC of weakly interacting
exciton-polaritons in semiconductor microcavities \cite{Kasprzak} and a  
non-equilibrium superradiant state using ultra-cold atoms in optical cavities \cite{Baumann2010}.
Theoretical interest has focused on a possible Mott insulator-superfluid transition of polaritons in coupled-cavity arrays as described
by the Jaynes-Cummings-Hubbard model (JCHM), where each cavity is strongly coupled to a two-level system (2LS) and photons
can hop between cavities \cite{Greentree2006,Hartmann2006,Angelakis2007, Rossini2007,Aichhorn2008,Schmidt2009,Koch2009,Schmidt2010_1}.
However, with a few exceptions \cite{Tomadin2010a, Hartmann2010, Wu, Nunnenkamp}, most of these studies did not take into account the basic nature of quantum optical applications, drive and dissipation.

In this paper, we study the coherently pumped JCHM including dissipation via spontaneous emission of the 2LS and cavity loss. We compare exact numerical
simulations of a dimer of two coupled cavities with mean-field theory for an extended array.
We calculate the pump frequency dependence of the photon field as measured in a homodyne/heterodyne detection scheme, 
the second-order coherence function (photon statistics), and the fluorescence (emission) spectra.
At weak hopping, we find strong signatures of photon blockade as observed in single-cavity systems. 
At strong hopping, the state of the photons in the driven-dissipative array depends on its size. 
In a dimer model consisting of two coupled cavities we find a photon blockade even at large hopping.
For a lattice consisting of infinitely many coupled cavities the blockade effect vanishes and a coherent photon state emerges,
 which can be described semi-classically.
We find that the crossover from weak to strong hopping is smooth as long as the array is pumped at the bottom of the lower polariton band.
Higher pump frequencies may cause tunneling-induced bistabilities.

The driven JCHM is described by the Hamiltonian
\begin{equation}
\label{jchm}
H = \sum_i h_i^\mathrm{JC} - \frac{J}{z}\sum_{\langle ij \rangle}a_i^\dagger a_j + f\sum_i \left(a_i^\dagger + a_i\right).
\end{equation}
Here, $h_i^\mathrm{JC}$ denotes the Jaynes-Cummings model (JCM) at site $i$ of the array,
\begin{equation}
\label{jcm}
h_i^\mathrm{JC} = \delta_r a_i^\dagger a_i + \frac{\delta_q}{2} \sigma_i^z + g\left( \sigma_i^+ a_i + \sigma_i^- a_i^\dagger\right)\,.\label{eq:hamiltonian}
\end{equation}
Each 2LS is represented by a spin operator and coupled to the cavity photons with strength $g$. 
The second term in (\ref{jchm}) describes the hopping of photons to one of its $z$ nearest neighbors at a rate $J/z$
(such that the total bandwidth is $2J$). All cavities are subject to a coherent laser drive of strength $f$ described 
by the third term in (\ref{jchm}). Working in a frame that rotates at the
pump (laser) frequency $\omega_l$, we measure the energies of
resonators (cavities) and 2LS (qubits) with respect to
the pump, $\delta_{r(q)}=\omega_{r(q)}-\omega_l$. 
Dissipation is taken into account in the master equation for
the density matrix
\begin{equation}
\label{master}
\dot{\rho}=-i \left[H,\rho\right]+(\kappa/2)\mathscr{L}[a_i] + (\gamma/2)\mathscr{L}[\sigma_i^-], \label{eq:neumann}
\end{equation}
with $\mathscr{L}[\sigma_i^-] = \sum_i
\left(2\sigma_i^-\rho\sigma_i^+-\sigma_i^+
  \sigma_i^-\rho-\rho\sigma_i^+ \sigma_i^-\right)$ and a similar decay
term for the photon operators. Here, $\kappa$ denotes the cavity decay rate and $\gamma$ the spontaneous emission rate of the 2LS.\\
\begin{figure}[t]
\centering
 \includegraphics[width=0.5\textwidth,clip]{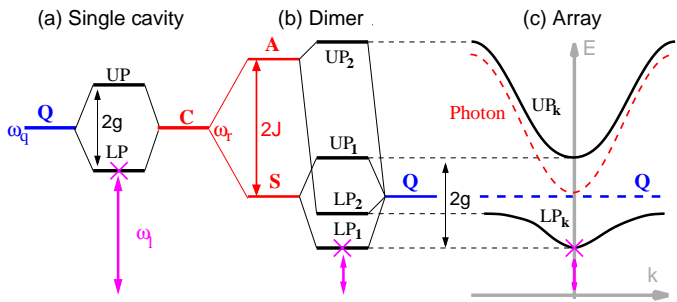}  
 \caption{Level scheme for the lowest excitations of the JCHM for (a) a single cavity, (b) a dimer model, (c) a coupled cavity array. Hopping between cavities gives (anti-)symmetric superpositions of photon states in (b) and photon bands $\epsilon({\bf k})=J\cos({\bf k})$ in (c).  The qubits Q are resonant with the lowest photon states: (a) the cavity mode C ($\omega_q=\omega_r$), (b) the symmetric superposition of photon states ($\omega_q=\omega_r-J$), (c) the bottom of the photon band ($\omega_q=\omega_r-J$). This gives rise to dressed (polariton) states (LP, UP) in (a) and (b) and polariton bands in (c). In this paper the laser frequency $\omega_l$ is near the lowest excitations in the system (bottom of the polariton band).}
\label{fig:levelscheme}
\end{figure}
%

\begin{figure}[t]
\centering
 \includegraphics[width=0.43\textwidth,clip]{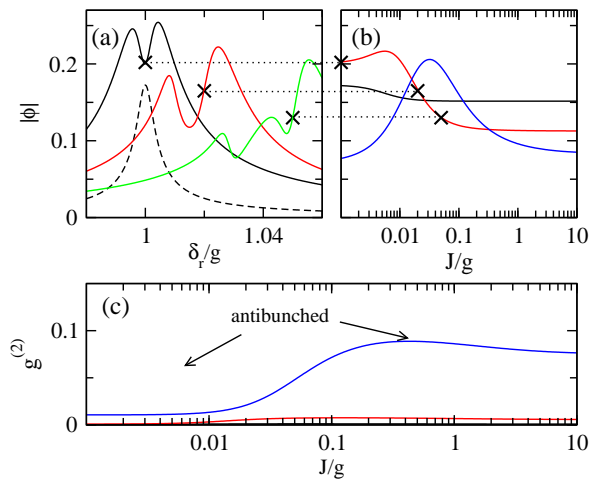}  
 \caption{Photon field $|\phi|$ and second-order coherence $g^{(2)}$ for the dimer of two coupled cavities, (a) as a function of resonator detuning $\delta_r/g$ at hopping $J/g=0\, ({\rm black}), 0.02\, ({\rm red}), 0.05\, ({\rm green})$ and drive strength $f/g=0.005$, and (b) as a function of hopping strength $J/g$ at drive $f/g=0.001\,({\rm black}), 0.005\,({\rm red}), 0.02\,({\rm blue})$
when the laser frequency is resonant with the lowest excitation ($\delta_r=g+J$, see Fig.~\ref{fig:levelscheme}). The dashed line in (a) corresponds to a single cavity ($J=0$) at very low drive $f/g=0.001$. The crosses mark points in (a) where $\delta_r=g+J$. Panel (c) shows $g^{(2)}$ as a function of the hopping strength $J/g$ for the same drive strengths as in (b). All dissipation rates are $\kappa=\gamma=0.005g$.}
\label{fig:dimer}
\end{figure}

Recently, it was shown that the single driven and dissipative JCM ($J=0$) for zero detuning ($\omega_r=\omega_q$) yields surprisingly complex
behavior  \cite{bishop-nphys, Bishop2010}, which we briefly summarize here.  
The local Hamiltonian in (\ref{jcm}) has eigenstates
that are symmetric and anti-symmetric superpositions of excited 2LS
and photon states, upper and lower polariton dressed
states (Fig.~\ref{fig:levelscheme}a).  
Pumping weakly near the lower polariton, $\delta_r \approx g$, yields a resonance
in the photon field $\phi=|\langle a
\rangle|$ with Lorentzian lineshape (dashed line in Fig.~\ref{fig:dimer}a).
However, beyond linear response the homodyne signal turns into anti-resonant behavior and the Lorentzian develops
a central dip of width $\sim f/\kappa$ (black line in Fig.~\ref{fig:dimer}a). This effect can be understood by restricting the
Hilbert space to two states, the vacuum and the state with a single
lower polariton (LP). The anti-resonance arises when
this effective 2LS saturates. This effect is the semi-classical Rabi splitting corresponding to the dressing of dressed states \cite{carmichael92} which has recently been observed in circuit QED as a Mollow triplet in fluorescence spectra \cite{wallraff}. 
The 2LS approximation is appropriate as the nonlinearity $U_{\rm eff}=\epsilon_{2}-2\epsilon_{1}=g(2-\sqrt{2})$ of the JCM prevents higher states from being excited ($\epsilon_{1,2}$ are the lowest energies in the Hilbert space sector with $1$ or $2$ excitations, respectively). 

We now consider extended systems as described by the JCHM in Eq.~(\ref{jchm}) with $J\neq 0$. 
The level scheme for the lowest excitations is shown in Fig.~\ref{fig:levelscheme},
where we choose the qubits to be resonant with the symmetric photon state of the dimer consisting of two
coupled cavities (Fig.~\ref{fig:levelscheme}b) or the bottom of the photon band in the infinite array (Fig.~\ref{fig:levelscheme}c).
At weak hopping we expect the coupled-cavity array to exhibit a similar blockade effect as a single cavity.
At strong hopping the effective nonlinearity in the spectrum to leading order in $g\ll J$ is $U_{\rm eff}=2g\left(1-\sqrt{1-1/(2N_s)}\right)$ for a 1D chain. This only vanishes as the number of lattice sites $N_s$ becomes large. Small arrays are thus expected to always show photon blockade, large ones only at small hopping.

Figure \ref{fig:dimer} shows the photon field $\phi=|\langle a_1
\rangle|$ and the second-order coherence $g^{(2)}=\langle a_1^\dagger a_1^\dagger a_1 a_1\rangle/\langle a_1^\dagger a_1\rangle^2$ for a dimer model consisting of two coupled cavities as
obtained from exact numerical evaluation of the master equation in (\ref{master}).
In Fig.~\ref{fig:dimer}a, the anti-resonance broadens when the hopping strength increases and shifts to larger values of detuning $\delta_r/g$, i.e., smaller pump frequencies.
The appearance of a second anti-resonance for $J=0.05g$ is associated with a two-polariton state with one polariton in each cavity.
A similar effect has been reported for a three-site ring lattice in \cite{Nunnenkamp}.
The right-hand panel in Fig.~\ref{fig:dimer} shows the crossover to the large hopping regime for several values of the pump strength. 
The anti-resonant feature remains for $J\gg g$, i.e., the photon field decreases as a function of pump strength at fixed hopping. 
The second-order coherence $g^{(2)}$ in Fig.~\ref{fig:dimer} confirms this photon blockade picture.
At weak drive ($f \ll U_{\rm eff}$) photons in the dimer remain anti-bunched with $g^{(2)}\ll 1$ even at strong hopping.
Note, that the dimer model is realizable with current state-of-the-art technology and has been shown to exhibit interesting 
new physics due to the coupling between cavities in various other contexts \cite{Schmidt2010_2, Leib2010, Knap2011a, Liew, Bamba,Ferretti10}.
\begin{figure}[t]
\centering
 \includegraphics[width=0.48\textwidth,clip]{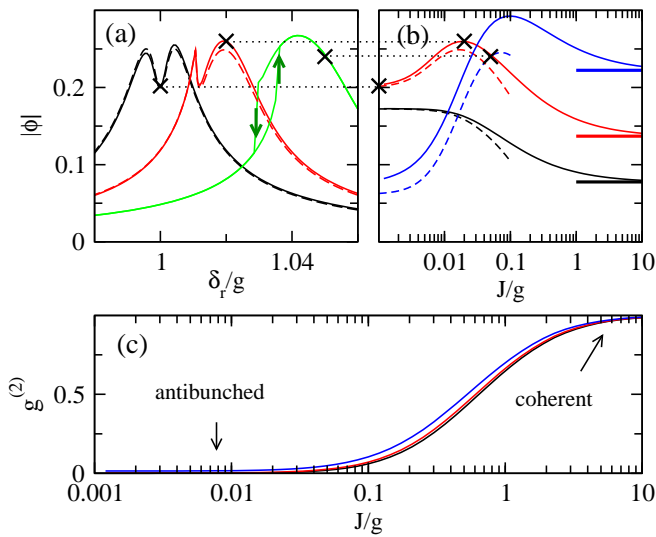}  
\caption{Photon field $|\phi|$ (upper panels) and second-order coherence $g^{(2)}$ (lower panel) for the array 
as obtained from mean-field theory for the same parameters as in Fig.~\ref{fig:dimer}.
Dashed lines in (a) and (b) correspond to the effective spin model in Eq.~(\ref{eq:heff}). The solid horizontal lines 
in (b) are the semiclassical asymptotes for Eq.~(\ref{dicke}).}
\label{fig:array}
\vspace{-0.4cm}
\end{figure}

We now increase the coordination number $z$ and consider a large array of coupled cavities. Continuous bands form between the upper and lower polariton states (Fig.~\ref{fig:levelscheme}c) and the full problem is then no longer numerically tractable. 
For large $z$, the hopping term is approximated by $a_i^\dagger a_j \approx \langle
a_i^\dagger\rangle a_j + a_i^\dagger \langle a_j \rangle - \langle
a_i^\dagger\rangle \langle a_j \rangle $. As all cavities are
equivalent, we take the photon expectation values to be the same
at each site of the array $\phi = \langle a_i\rangle$. This is the same mean-field
decoupling as in the corresponding equilibrium
model \cite{Greentree2006}. It was first used out-of-equilibrium to calculate the dynamics of the dissipative Bose-Hubbard model \cite{Tomadin2010a}.
One may then find the steady state of this decoupled on-site problem numerically (Fig.~\ref{fig:array}).
As in the dimer model, the anti-resonance shifts and broadens with increasing hopping. 
The two sidepeaks of the anti-resonance become asymmetric and turn into a jump corresponding to a hopping-induced bistability. 
Similar bistable behavior in quantum optics is well known for a single strongly driven cavity \cite{Lugiato84,Alsing91,Bishop10,Savage88} and for the Dicke model \cite{Bowden79}.  
Here, a bistability develops at fixed pump strength when increasing the hopping rate $J$ between cavities. Note, that these instabilities only appear when the array is pumped above the bottom
of the lower polariton band. In Fig.~\ref{fig:array}b we show the crossover to large hopping. We pump at the bottom of the polariton band, where no instabilities occur.
The curves for different pump strengths cross and reorder, such that at strong hopping the photon field increases
uniformly with drive strength. Thus, the anti-resonance feature vanishes. 
The second-order coherence function $g^{(2)}$ undergoes a smooth crossover from anti-bunching at weak hopping with $g^{(2)}\approx 0$ to a coherent photon state at strong hopping
with $g^{(2)}\approx 1$ even at vanishingly small drive strength. Thus, all signatures of the photon blockade effect are destroyed in a large array at strong hopping.\\
%
%

%
\begin{figure}[t]
\centering
\includegraphics[width=0.45\textwidth,clip]{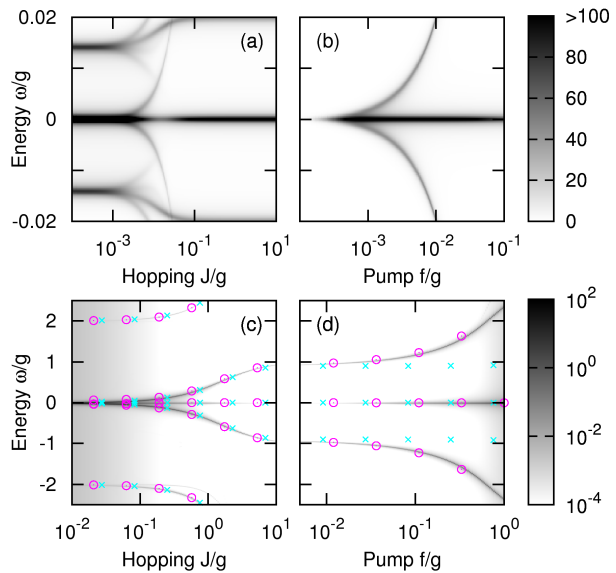}  
\caption{Fluorescence spectra $S(\omega)$ for the dimer (upper panels) and for the array (lower panels) at fixed drive strength (a) $f=0.01g$, (c) $f=0.005g$ and fixed hopping (b,\,d) $J=10g$. The laser frequency is resonant with the lowest polariton state ($\delta_r=g+J$, see Fig.~\ref{fig:levelscheme}). Blue crosses and red circles in (c,\,d) correspond to the effective models at weak drive in Eq.~(\ref{eq:heff}) and strong hopping in Eq.~(\ref{dicke}), respectively. The decay rates are $\kappa=\gamma=0.001g$ in (a,\,b) and $\kappa=\gamma=0.005g$ in (c,\,d).}
\label{fig:rf}
\end{figure}
Another important experimental signature is the fluorescence spectrum.
We calculate the emission spectrum as the
Fourier transform $S(\omega)$ of the on-site auto-correlation function $S(\tau) =
\langle a^\dagger(t+\tau) a(t) \rangle$ in the steady state of the system.
Results are shown in Fig.~\ref{fig:rf}.
%
The spectrum of the dimer (Fig.~\ref{fig:rf}) exhibits a 
Mollow triplet of emission peaks at weak hopping at $\omega\approx 0,\pm\sqrt{2}f$ 
arising from transitions between the vacuum and the lower polariton state (${\rm LP}_1$) dressed by the laser field.
This agrees with recent circuit QED experiments in a single cavity
and can be interpreted as a signature of photon blockade \cite{wallraff}. 
For large hopping, the low-energy spectrum again exhibits a Mollow triplet,
however with modified frequencies near $\omega\approx \pm 2f$, which is due to a shift of
the weight of the photonic component in the lower polariton state ${\rm LP}_1$ with increasing hopping.
%

The lower panel in Fig.~\ref{fig:rf} shows the fluorescence spectra for the array as obtained from mean-field theory.
At small $J$, the lower polariton mode at $\omega\approx 0$  splits and the corresponding side-peaks move to higher frequencies
as hopping increases. Despite the absence of the photon blockade the triplet structure survives at large hopping. This is because the resonant frequency of a single cavity is different from that of the array --- it is detuned by $J+g$ from the pump. In the limit of small drive $f$, where the photon field vanishes, the emission spectrum is hence due to off-resonance fluorescence of pumping a single 2LS at $g$ below its transition frequency, giving peaks at $\pm g$ (Fig. \ref{fig:rf}d). Additional peaks with vanishing spectral weight as $J\rightarrow 0$ arise at higher energies $\omega\approx\pm2g$ due to the emission from upper polariton states (for the dimer these modes are not visible on the scale shown).\\

Further insight into the behavior of the array at weak driving ($f \ll U_{\rm eff}$, where $U_{\rm eff}\to 0$ when $J\gg g$) can be gained through an effective model
by generalizing the two-level dressed state approximation derived for a single cavity in \cite{bishop-nphys} to the JCHM. By projecting into this restricted Hilbert space, we obtain an effective Hamiltonian 
\begin{eqnarray}
\label{effmodel_lowJ}
    H_\mathrm{eff}=\frac{\tilde{J}}{z}\sum_{\langle
      ij\rangle}\tau_i^+\tau_j^- + \frac{\epsilon}{2}\sum_i \tau_i^z +
    \tilde{f}\sum_i\left(\tau_i^+ + \tau_i^-\right)\,,\label{eq:heff}
\end{eqnarray}
which is valid when the number of excitations per
cavity is small.
The Pauli matrices $\{\tau^+,\tau^-,\tau^z\}$ describe the
transition between the vacuum and the lower polariton at energy
$\epsilon = \delta_r\sin^2\theta+\delta_q\cos^2\theta$ 
with $\theta=\tan^{-1}(-2g/(\delta_r-\delta_q))/2$.
Here, $\tilde{J}=J\sin^2\theta$ and
  $\tilde{f}=f\sin(\theta)$ are
effective hopping and pump strengths.
The master equation for the effective system becomes
$\dot{\rho}=-i[H_{\text{eff}},\rho]+\tilde{\kappa}\,\mathscr{L}[\tau^-]$
with the decay rate
$\tilde{\kappa}=\kappa\sin^2\theta+\gamma\cos^2\theta$.
This model is equivalent to a dissipative Heisenberg $XX$ model in a magnetic field. 
Dissipative spin chains
\cite{Prosen2009,Prosen2010,Znidaric2010}, with dissipation acting
only on the boundary sites, have recently been considered as
examples of non-equilibrium quantum phase transitions.  The current
problem, however, differs from these in having dissipation on all
sites.  Dissipative spin chain problems also arise in the context of the Rydberg blockade \cite{Lee2011,Lee2012}. 
%
Using the same mean-field approximation as above produces an effective 2LS Hamiltonian $H_{\rm eff}=(\epsilon/2) \tau_z+(f_{\rm eff}\tau^++{\rm h.c.})$
with an effective drive strength $f_{\rm eff}=\tilde{J}\psi+\tilde{f}$ and the self-consistent field $\psi=\langle \tau^-\rangle$.
The dashed lines in Fig.~\ref{fig:array} correspond to the steady state of this model and agree well with the numerical treatment at small drive and hopping. Since $U_\mathrm{eff}\rightarrow 0$ for large $J$, the blockade picture breaks down with increasing hopping. The emission spectra resulting from this approximation are shown as blue crosses in Fig.~\ref{fig:rf}. \\

%
%
%
The strong hopping regime of the full lattice can be understood by relating the JCHM  to a pumped dissipative semi-classical model.  Rather than restricting the Hilbert space to low excitation numbers, we let the photons occupy a coherent state and factorize products of spin and photon expectation values. The on-site equations of motion are
\begin{eqnarray}
\label{dicke}
\dot{\phi}_i&=& -i(\delta_r \phi_i + f + g s^-_i  - (J/z)\sum_{\langle ij \rangle}\phi_j) - (\kappa/2)\phi_i \nonumber\\
\dot{s}_i^-&=& -i \delta_q s^-_i + 2 i g \phi_i s^z_i - (\gamma/2) s^-_i \label{eq:sc2}\\
\dot{s}_i^z&=& -i g (\phi_i {s^-_i}^* - \phi_i^* s^-_i )-\gamma\left(s^z_i+1/2\right)\nonumber\,,
\end{eqnarray}
where $s_i^\pm=\langle \sigma_i^\pm\rangle$, $s_i^z=\langle \sigma_i^z\rangle$ and $\phi_i=\langle a_i\rangle$. 
In the steady state we assume that all sites are identical, $\phi_j=\phi_i$.
Pumping at the bottom of the polariton band (Fig.~\ref{fig:levelscheme}c) we have $\delta_r=g+J$, such that the
dependence on hopping disappears for the steady state in (\ref{dicke}). 
Fig. \ref{fig:array}b shows that the corresponding asymptotes (solid horizontal lines) are indeed approached by the mean-field numerics for $J\rightarrow\infty$. 
The low-energy spectra $S(\omega)$ in this approximation are found by linearizing equations (\ref{eq:sc2}) about their steady state. Fluctuations on different sites are uncorrelated in (\ref{dicke}) and the $1/z$ term can thus be neglected. This introduces a $J$-dependence into the spectrum. Fig.~\ref{fig:rf} shows that the predictions of this model match the full numerics of the lattice mean-field theory everywhere except at very small hopping (deviations not visible on the scale shown).
The equations of motion in (\ref{dicke}) are the same
as those obtained from a Dicke model where the single site expectation values in (\ref{dicke}) are replaced by collective spin variables. However, the spectra differ from the usual Dicke model in describing on-site rather than collective fluctuations.\\

In summary, we have investigated the crossover from weak to strong hopping in the coherently driven dissipative JCHM. In the weak hopping limit, blockade physics survives
and both the extended and two-site arrays behave similarly. At large hopping, blockade effects survive in the two-site system, while
the infinite system becomes semiclassical and can show bistability for certain pump frequencies. Our work thus provides a strong motivation for further experimental and theoretical studies of coupled cavity arrays.

\begin{acknowledgments}
  We are grateful to A. Wallraff and J. Fink for discussions.  J.K.
  acknowledges funding from EPSRC grants EP/G004714/2 and
  EP/I031014/1. H.T. acknowledges support from the Princeton Center for Complex Materials under grant no. DMR-0819860 and from the Swiss NSF through grant no. PP00P2-123519/1. S.S. acknowledges support from the Swiss SNF through the NCCR MaNEP.
\end{acknowledgments}

%


%

\end{document}